\documentclass[aps,prd,twocolumn,superscriptaddress]{revtex4-2}
\usepackage{epsfig}
\usepackage{amsmath}
\usepackage{amsthm}
\usepackage{amsfonts}
\usepackage{amssymb}
\usepackage{dsfont}
\usepackage{multirow}
\usepackage{appendix}
\usepackage{slashed}
\usepackage[active]{srcltx}
\usepackage{psfrag}
\usepackage[colorlinks,
linkcolor = green,
urlcolor  = blue,
citecolor = blue,
anchorcolor = blue]{hyperref}
\begin{document}
	\title{Scalar fully-charm and bottom tetraquarks under extreme temperatures}
	\date{\today}
	\author{A.~Aydın}
	\affiliation{Department of Physics, Kocaeli University, 41001 Izmit, T\"urkiye}
	\author{H.~Sundu}
	\affiliation{Department of Physics Engineering, Istanbul Medeniyet University, 34700 Istanbul, T\"urkiye}
	\author{J.Y.~S\"ung\"u}
	\affiliation{Department of Physics, Kocaeli University, 41001 Izmit, T\"urkiye}
	\author{E. Veli Veliev}
	\affiliation{Department of Physics, Kocaeli University, 41001 Izmit, T\"urkiye}
	\begin{abstract}
		Temperature-dependences of masses and current couplings of the ground state of the fully heavy tetraquarks $T_{4c}$ and $T_{4b}$, composed of charm (c) and bottom (b) quarks and antiquarks with spin-parities $J^{PC} = 0^{++}$ are evaluated in the diquark–antidiquark picture using Thermal QCDSR including vacuum condensates up to dimension four. The calculated values for $cc\bar{c}\bar{c}$ and $bb\bar{b}\bar{b}$ tetraquark states at $T=0$ align well with the experimental data on the broad structures. Based on the numerical analyses around the critical temperature, the mass of the $T_{4c}$ state decreases by $8\%$ compared to its vacuum state, while for its b partner, this percentage is approximately $3.3\%$. For the decay constants, the reductions are approximately $71\%$ and $66.6\%$, respectively. The precise exploration of tetraquark states awaits future scrutiny in upcoming experiments, including Belle II, Super-B, PANDA, and LHCb.
\end{abstract}
\maketitle
\section{Introduction}
Theoretically predicted new state of matter Quark-Gluon Plasma (QGP) is believed to have existed in the first microseconds after the Big Bang. QGP emerges under extreme conditions of high temperature and density, allowing quarks and gluons, the fundamental building blocks of protons and neutrons, to roam freely – a phenomenon known as deconfinement. The significance of QGP research lies in its potential to validate theories of the strong force (Quantum Chromodynamics), offer insights into the early universe, and shed light on the confinement mechanism that binds quarks within nucleons. Exploration of QGP paves the way for a deeper understanding of the fundamental forces and particles that govern our universe~\cite{Shuryak:2008eq}.

The QGP is achieved in particle accelerators such as the Large Hadron Collider (LHC) and the Relativistic Heavy Ion Collider (RHIC) by colliding heavy ions, like gold or lead, at near-light speeds. These collisions generate extreme temperatures and energy densities, allowing quarks and gluons to move freely for a brief moment. Studying QGP helps scientists understand the strong force that binds quarks and provides insights into the early universe's conditions~\cite{Ayala:2020rmb,Ayala:2016vnt,Steinbrecher:2018phh,Fischer:2018sdj}. 

Key experiments by collaborations such as ALICE at the LHC~\cite{ALICE:2008ngc} in central Pb-Pb collisions and STAR at RHIC~\cite{STAR:2005gfr} have been instrumental in probing the properties of QGP in central Au+Au  collisions, contributing to our understanding of quantum chromodynamics (QCD) and the fundamental structure of matter. Several new research facilities and experimental setups have emerged to push forward the study of QGP \cite{CMS:2024krd,STAR:2024bpc,Wang:2023xhn}. Investigations into various reactions such as Ar+Sc, Xe+La, and Pb+Pb are being actively pursued, focusing on hadron spectra and fluctuations \cite{Podlaski:2024kxg}.

Besides, exotic mesons, a fascinating class of particles in the realm of particle physics, defy the conventional quark-antiquark composition of traditional mesons~\cite{Brambilla:2019esw,Chen:2016qju,Agaev:TJP}. Unlike the standard quark model, which predicts mesons as bound states of a quark and an antiquark, exotic mesons possess unique internal structures that challenge our understanding of fundamental particle interactions. These exotic states may consist of more than a quark-antiquark pair and can include additional quarks, antiquarks, or gluons in complex configurations. Examples of exotic mesons include tetraquarks, which contain two quarks and two antiquarks, and glueballs, which are composed solely of gluons. The study of exotic mesons provides valuable insights into the strong force that governs the behavior of quarks and gluons within hadrons. 

Experimental searches for exotic mesons continue to push the boundaries of particle physics, shedding light on the rich spectrum of particles and interactions that exist in the universe. Moreover, many attempts are made to search for the effect of nuclear and medium on exotic meson properties~\cite{Azizi:2020itk,Azizi:2020yhs,Sungu:2024oax,
Veliev:2014tca,Veliev:2011kq,Sungu:2020zvk,Sungu:2019ybf,Turkan:2019anj,Gungor:2023ksu,Sungu:2020azn}. Recent observations at the Large Hadron Collider (LHC) have shed light on new, short-lived particles composed of charmed quarks~\cite{liu:2020eha}. In 2020, the LHCb collaboration made a significant announcement regarding the discovery of two intriguing resonances: one being a narrow structure identified as \( X(6900) \), and the other a broader structure observed within a defined mass range. These findings spurred further investigation by other LHC collaborations~\cite{LHCb:2020bwg}. Building on this initial discovery, subsequent experiments by CMS and ATLAS confirmed the existence of the $ X(6900) $ resonance. Moreover, these experiments unveiled two additional resonances, $ X(6600) $ and $ X(7300 $). These new findings highlight the ongoing exploration of exotic particles at the LHC and contribute to our understanding of the strong force that binds quarks together. The ATLAS collaboration investigated the di-$ J/\psi $ invariant mass spectrum and observed two interfering resonances, whose masses and widths were measured to be:
\begin{eqnarray}
X(6200):M = 6.22 \pm 0.05^{+0.05}_{-0.04} \mathrm{GeV},\notag\\ 
\Gamma = 0.31 \pm 0.12^{+0.07}_{-0.08}  \mathrm{GeV},\notag\\ 
\end{eqnarray}
\begin{eqnarray}
	X(6600): M = 6.62 \pm 0.03^{+0.02}_{-0.01} \mathrm{GeV},\notag\\
	\Gamma = 0.31 \pm 0.09^{+0.06}_{-0.11} \mathrm{GeV}.
\end{eqnarray}
Recently, a new resonance, \( X(6200) \), was observed in the CMS experiment within the \( J/\psi J/\psi \) mass spectrum from proton-proton collisions at a center-of-mass energy of \( \sqrt{s} = 13 \) TeV \cite{CMS:2023owd}. This observation reaches a local significance exceeding \( 5 \) standard deviations, with the resonance located at a mass of 
\begin{eqnarray}
X(6600):M=6638^{+38}_{-43}(\text{stat})^{+31}_{-16}(\text{syst}) \, \text{MeV}.
\end{eqnarray}
Of the two new structures seen now by CMS in the $ J/\psi J/\psi $ mass spectrum, only the one at 6600 MeV has a significance that exceeds the famous 5 sigma threshold. Another structure of notable significance has been identified with a mass of 
\begin{eqnarray}
X(6900): M=6847^{+28}_{-44}(\text{stat})^{+20}_{-48}(\text{syst}) \, \text{MeV},
\end{eqnarray}
which corresponds to the $X(6900)$ resonance previously reported by the LHCb experiment and confirmed by the ATLAS experiment.

Furthermore, evidence for an additional new resonance, observed with a local significance of $4.7$ standard deviations, has been found at a mass of 
\begin{eqnarray}
X(7200): M=7134^{+48}_{-25}(\text{stat})^{+15}_{-41}(\text{syst}) \, \text{MeV}.
\end{eqnarray}

The results also include a model without interference, which shows a poorer fit to the data and reveals mass shifts of up to 150 MeV compared to the model with interference. This format ensures clarity by highlighting the specific details of each observed structure concisely.

As for theoretical studies, Ref.~\cite{liu:2020eha} confirms that the $ X(6600) $ structure observed in the di-$ J/\psi $ invariant mass spectrum can indeed be explained by the 1S-wave state $ T_{(4c)}(6550)$. This interpretation aligns with theoretical models predicting the existence of fully charmed tetraquarks. The observed mass and properties of the $ X(6600) $ fit well with this 1S-wave state, providing a coherent explanation for the experimental data. Also, Agaev et al. derived the mass of $ X(6600)$ in the QCSR framework nicely agrees with experimental data as $ m=(6570 \pm 55) $ MeV and its b-partner $ m = (18540\pm 50) $ MeV~\cite{Agaev:2023wua}. The $ X(6200)$ observed by ATLAS is very close
to Faustov et al. prediction for the lowest ground state $ 0^{++} $ with the mass $ 6190 $ MeV. According to their study $ X(6200), X(6400), X(6600), X(7200)$, and $X(7300)$
can also be interpreted as the different excitations of the fully charmed tetraquarks~\cite{Faustov:2022mvs}. Also, Lin and et al's calculations suggest that the observed $ X(6600), X(6900), $ and $ X(7300) $ likely represent distinct radial excitations of the $ [cc][\bar{c}\bar{c}] $ system~\cite{Lin:2024olg}. According to Ref.~\cite{Chen:2016jxd}, based on QCD sum rule analysis, their findings indicate that the broad structure spanning $ 6.2-6.8$ GeV could represent an S-wave \( cc\overline{c}\overline{c} \) tetraquark state with \( J^{PC} = 0^{++}, 2^{++} \) configurations, while the narrow structure around $ 6.9 $ GeV may correspond to a P-wave state with \( J^{PC} = 0^{+}, 1^{+} \). In Ref.~\cite{Wang:2022xja} Z. Wang's predictions suggest the following assignments for the tetraquark states: the $X(6220)$ is proposed to be the ground state tetraquark with $J^{PC} = 0^{++}$ or $1^{+-}$. The $X(6620/6600)$ is likely the first radial excited tetraquark state with $J^{PC} = 0^{++}$ or $1^{+-}$. The $X(6900)$ is identified as the second radial excited tetraquark state with $J^{PC} = 0^{++}$. Lastly, the $X(7220/7300)$ is assigned as the third radial excited tetraquark state with $J^{PC}= 0^{++}$. Another study proposes that the $ X(6600) $ and $ X(6400) $ states are the tensor $ (2^{++}) $ and scalar $ (0^{++}) $ states of an S-wave multiplet of $ cc\bar{c}\bar{c} $ tetraquark states~\cite{Anwar:2023fbp}. Also, in Ref.~\cite{Yan:2023lvm} they find that the fully charmed tetraquark $ cc\bar{c}\bar{c} $ with $J^{PC} = 0^{++}$ has mass about
$ 6572 $ MeV and $ 19685 $ MeV for the b-partner. Yang et al.'s findings indicate that the broad structure observed between $ 6.2 $ and $ 6.8 $ GeV can be attributed to the $0^{++}$ octet-octet tetraquark states, with estimated masses of $6.44 \pm 0.11$ GeV and $6.52 \pm 0.10$ GeV. Furthermore, when extending this analysis to the b-quark sector, the corresponding fully-bottom tetraquark partners are predicted to have masses in the range of approximately $ 18.38  $ to $ 18.59 $ GeV~\cite{Yang:2020wkh}. The masses of the low-lying 1P \( cc\bar{c}\bar{c} \) system with \( J^{PC} = 0^{-+}, 1^{-+}, \) and \( 2^{-+} \) are predicted to be $ 6666 $ MeV, $ 6624 $ MeV, and $ 6647 $ MeV, respectively. Another tetraquark state, also named $ X(6600) $, was reported with a mass of \( 6620 \pm 10 \) MeV by the ATLAS Collaboration, suggesting it may be a mixture of these low-lying 1P states in Ref.~\cite{Yu:2022lak}.

Fully heavy quark systems play a crucial role in studying the properties of QGP. Due to their significant mass, these heavy quarks have a longer lifespan compared to their lighter counterparts. This extended existence allows them to escape the fleeting QGP before decaying, carrying valuable information about the medium they interacted with. Additionally, their hefty mass makes them less likely to participate in hadronization within the short QGP lifetime. This minimal hadronization preserves their identity, making them clean and reliable probes of the QGP. By studying the momentum distribution, energy loss, and thermalization of these heavy quarks, scientists can infer the transport properties, strength of the strong force, and coupling between the QGP and these heavy particles. In essence, fully heavy quarks act as essential messengers, delivering crucial data that help us unravel the mysteries of the QGP~\cite{Brambilla:2010cs}.

This article investigates the temperature dependence of the mass and coupling constant of \( X(6600) \) and its \( b \) partner at finite temperatures within the framework of thermal QCDSR.
Such investigations are important for validating and interpreting heavy-ion collision experiments and can provide insights into the QGP. This is because these particles form within the QGP medium.

The paper is structured as follows. Section \ref{sec:two} provides a summary of the Thermal QCD sum rules (TQCDSR) approach utilized in our calculations. Section \ref{sec:Num} analyzes the mass and current coupling constant of $X(6600)$, denoted as $ T_{4c} $ and its b-partner $ T_{4b}$. 
%
\section{FORMALISM}\label{sec:two}
The QCD Sum Rules (QCDSR) model, also known as the QCD sum rule method, is a theoretical framework within quantum chromodynamics (QCD) used to study the properties of hadrons, particularly their masses, decay constants, and other parameters. It combines elements of quantum field theory, perturbation theory, and phenomenology to extract information about hadronic properties from the fundamental theory of strong interactions, QCD. The basic idea behind QCDSR is to relate the properties of hadrons to vacuum expectation values of quark and gluon operators, which can be computed theoretically. These vacuum expectation values are encapsulated in quantities known as condensates. By constructing correlation functions involving quark and gluon fields, one can derive sum rules that relate these correlation functions to the desired hadronic properties \cite{Shifman:1978bx,Reinders:1984sr}. To study how the mass and current coupling of the $T_{4c}$ state vary with temperature, we apply the QCDSR formalism in the context of TQCDSR. The computation initiates by outlining the thermal correlator~\cite{Bochkarev:1985ex}:
\begin{equation}\label{eq:CorrF1}
	\Pi(q,T)=i\int d^{4}x~e^{iq\cdot x}\langle\Phi|
	\mathcal{T}\{\eta(x) \eta^{\dag}(0)\}|\Phi\rangle,
\end{equation}
here $\Phi$ represents the thermal medium, $\mathcal{T}$ is the time-ordered operator, $\eta(x)$ is the interpolating current associated with the $T_{4c}$ resonance, and $T$ shows the temperature.

TQCDSR is an extension of the conventional QCDSR to finite temperature medium. In standard QCDSR, properties of hadrons are studied in the vacuum, but at finite temperatures, such as those encountered in heavy-ion collisions or the early universe, the properties of hadrons can change due to the effects of temperature and density. TQCDSR provides a framework to investigate these changes by incorporating thermal effects into the calculations~\cite{Azizi:2019cmj}. This involves modifications to the correlation function and spectral densities to account for the thermal medium. By applying the operator product expansion (OPE) technique, TQCDSR can extract information about hadron properties at finite temperatures, such as masses, decay constants, decay widths, and form factors.

One of the key challenges in TQCDSR is the treatment of thermal effects on quark and gluon condensates, which are crucial parameters in QCDSR. Various approaches have been proposed to address this issue, including the use of temperature-dependent condensates and the inclusion of new operators in the OPE to capture thermal contributions~\cite{Mallik:1997pq,Furnstahl:1992pi,Dominguez:2010mx,Gubler:2011ua}.

To derive the TQCDSR, we begin by evaluating the correlation function using the physical degrees of freedom. Saturating this function with a complete set of states having the same quantum number $ J^{P}=0^{++} $ as the $T_{4c}$ state, and integrating Eq.~(\ref{eq:CorrF1}) over $x$, we get:
\begin{equation}
	\Pi^{\mathrm{Had.}}(q^2,T)=\frac{\langle
		\Phi|\eta|T_{4c}(q)\rangle\langle T_{4c}(q)|\eta^{\dagger
		}|\Phi\rangle}{m_{T_{4c}}^{2}(T)-q^{2}}+ \cdots,
\end{equation}
where \( m_{T_{4c}}(T) \) denotes the mass of the \( T_{4c} \) ground state as a function of temperature, with the ellipsis indicating the contributions from higher excited states and the continuum. The temperature-dependent current coupling constant is defined by the matrix element:
\begin{equation}\label{eq:Res1}
	\langle\Phi|\eta|T_{4c}(q)\rangle=f_{T_{4c}}(T)m_{T_{4c}}(T).
\end{equation}
Consequently, the physical representation of the correlation function, involving the thermal ground state mass and the current coupling constant, is formulated as:
\begin{equation}
	\Pi^\mathrm{Had.}(q^2,T) = \frac{f^2_{T_{4c}}(T) m^2_{T_{4c}}(T)}{m^2_{T_{4c}}(T) - q^2(T)} + \cdots. \label{eq:Phen2}
\end{equation}
The Eq.~(\ref{eq:Phen2}) is the invariant amplitude representing the hadronic part.

By isolating the ground state contributions from those of higher resonances and the continuum and then applying the Borel transformation, we express the hadronic side as:
\begin{eqnarray}\label{eq:CorBor}
	\mathcal{\widehat{B}}(q^{2})\Pi^{\mathrm{Had.}}(q^2,T)= m_{T_{4c}}^{2}(T)f_{T_{4c}}^{2}(T)~e^{-m_{T_{4c}}^{2}(T)/M^{2}},
\end{eqnarray}
where \( M \) denotes the Borel parameter in the QCDSR model.

Following this, we address the QCD component, where the correlation function is expressed in terms of quark and gluon degrees of freedom. The interpolating current for a tetraquark state with \( J^{PC}=0^{++} \) is defined by:

\begin{equation}
	\eta(x) = Q_a^T(x) C \gamma_\mu Q_b(x) \overline{Q}_a(x) \gamma^\mu C \overline{Q}_b^T(x).  \label{eq:CR1}
\end{equation}

In this expression, $a$ and $b$ are color indices, $Q(x)$ represents the charm ($c$) or bottom ($b$) quark fields, and $C$ is the charge conjugation matrix. This current construction ensures the correct symmetries for a tetraquark with the desired spin and parity.

The theoretical side of the correlator $\Pi_{\mu \nu }^{ \mathrm{Theor.}}(q,T)$ is encapsulated through a dispersion integral:

\begin{equation}
	\Pi^{ \mathrm{Theor.}}(q^2,T)=\int_{\mathcal{M}^2}^{\infty}\frac{\rho
		^{\mathrm{Theor.}}(s,T)}{s-q^{2}-i\epsilon}ds,
\end{equation}
where $\mathcal{M}^2=(4m_{c})^2$, and the spectral density function $\rho^{ \mathrm{Theor.}}(s,T)$ is the imaginary part of the correlation function:

\begin{equation}\label{eq:rhoQCD}
	\rho^{ \mathrm{Theor.}}(s,T)=\frac{1}{\pi}Im[\Pi^{\mathrm{Theor.}}].
\end{equation}

After extensive calculations, the  theoretical part of the correlation function, in terms of the heavy quark propagators, is written as:
\begin{eqnarray}
	\Pi^{ \mathrm{Theor.}}(q) &=& i\int d^{4}x e^{iq\cdot x} \{ \mathrm{Tr}[ \gamma_{\mu} \widetilde{\mathcal{D}}_{Q}^{b^{\prime}b}(-x) \gamma_{\nu} \mathcal{D}_{Q}^{a^{\prime}a}(-x)] \notag \\
	& \times& [\mathrm{Tr}[ \gamma^{\nu} \widetilde{\mathcal{D}}_{Q}^{aa^{\prime}}(x) \gamma^{\mu} \mathcal{D}_{Q}^{bb^{\prime}}(x)] \notag \\
	& -& \mathrm{Tr}[ \gamma^{\nu} \widetilde{\mathcal{D}}_{Q}^{ba^{\prime}}(x) \gamma^{\mu} \mathcal{D}_{Q}^{ab^{\prime}}(x)]] \notag \\
	& +& \mathrm{Tr}[ \gamma_{\mu} \widetilde{\mathcal{D}}_{Q}^{a^{\prime}b}(-x) \gamma_{\nu} \mathcal{D}_{Q}^{b^{\prime}a}(-x)] \notag \\
	& \times& \left[\mathrm{Tr}[ \gamma^{\nu} \widetilde{\mathcal{D}}_{Q}^{ba^{\prime}}(x) \gamma^{\mu} \mathcal{D}_{Q}^{ab^{\prime}}(x)] \right. \notag \\
	& -& \left.\mathrm{Tr}[ \gamma^{\nu} \widetilde{\mathcal{D}}_{Q}^{aa^{\prime}}(x) \gamma^{\mu} \mathcal{D}_{Q}^{bb^{\prime}}(x)] \right] \}, 
	\label{eq:QCD1}
\end{eqnarray}
where
\begin{equation}
	\widetilde{\mathcal{D}}_{c}(x) = C \mathcal{D}_{c}^{T}(x) C, \notag \\\label{eq:Prop}
\end{equation}
with \( \mathcal{D}_{Q}(x) \) being the \( Q\)-quark propagator.

The heavy quark propagator \(\mathcal{D}_{Q}^{ij}(x)\) for \( Q=c,b \) can be written according to Ref.~\cite{Reinders:1984sr}:

\begin{eqnarray}\label{eq:HeavyProp}
	&&\mathcal{D}_{Q}^{ij}(x)=i\int \frac{d^{4}k}{(2\pi )^{4}}e^{-ik\cdot x}\Bigg[ \frac{%
		\delta _{ij}\Big( {\!\not\!{k}}+m_{Q}\Big)
	}{k^{2}-m_{Q}^{2}}\nonumber \\
	&&-\frac{gG_{ij}^{\alpha \beta }}{4}\frac{\sigma _{\alpha \beta }\Big( {%
			\!\not\!{k}}+m_{Q}\Big) +\Big(
		{\!\not\!{k}}+m_{Q}\Big)\sigma_{\alpha
			\beta }}{(k^{2}-m_{Q}^{2})^{2}}\nonumber \\
	&&+\frac{g^{2}}{12}G_{\alpha \beta }^{A}G_{A}^{\alpha \beta
	}\delta_{ij}m_{Q}\frac{k^{2}+m_{Q}{\!\not\!{k}}}{(k^{2}-m_{Q}^{2})^{4}}+\cdots\Bigg],
\end{eqnarray}
where the shorthand notation for the external gluon field \(G_{ij}^{\alpha \beta}\) is used:
\begin{equation*}
	G_{ij}^{\alpha \beta }\equiv G_{A}^{\alpha \beta}\lambda_{ij}^{A}/2,
\end{equation*}
with \(\lambda_{A}^{ij}\) being Gell-Mann matrices, \(i,\,j\) as color indices, and \(A=1,\,2\,\ldots 8\) representing gluon flavors. The first term in Eq.~(\ref{eq:HeavyProp}) is the perturbative contribution to the heavy quark propagator, while the others are non-perturbative terms.

At finite temperatures, additional operators appear in the short-distance expansion due to the breaking of Lorentz invariance by the preferred reference frame, affecting the thermal heavy quark propagators~\cite{Azizi:2016ddw}. Additionally, we modify the vacuum condensates by their thermal averages. The gluon condensate, related to the gluonic part of the energy-momentum tensor $\theta _{\lambda \sigma }^{g}$, is given by \cite{Mallik:1997pq}:
\begin{eqnarray}\label{TrGG}
	&&\langle Tr^{c}G_{\alpha \beta }G_{\mu \nu }\rangle
	=\frac{1}{24}(g_{\alpha \mu }g_{\beta \nu }-g_{\alpha \nu
	}g_{\beta \mu })\langle G_{\lambda \sigma
	}^{a}G^{a\lambda \sigma }\rangle   \notag   \\
	&&+\frac{1}{6}\Big[g_{\alpha \mu }g_{\beta \nu }-g_{\alpha \nu
	}g_{\beta \mu }-2(u_{\alpha }u_{\mu }g_{\beta \nu }-u_{\alpha
	}u_{\nu }g_{\beta \mu }\notag \\
	&&-u_{\beta }u_{\mu }g_{\alpha \nu }+u_{\beta }u_{\nu }g_{\alpha \mu })\Big]%
	\langle u^{\lambda }{\theta }_{\lambda \sigma }^{g}u^{\sigma
	}\rangle.
\end{eqnarray}
Meanwhile, by substituting Eq.~(\ref{eq:HeavyProp}) and Eq.~(\ref{TrGG}) into Eq.~(\ref{eq:QCD1}), performing the Borel transformation for the $\Pi (M^{2},T)=\widehat{\mathcal{B}}\Pi^{Theor.} (q^{2},T)$ and applying standard procedures, we obtain the following expression as:
\begin{eqnarray}
\widehat{\mathcal{B}}\Pi^{Theor.} (q^{2},T)&=&\int_{\mathcal{M}^{2}}^{s(T)}ds\rho ^{\mathrm{Theor.}
	}(s,T)e^{-s/M^{2}}\notag   \\
&+&\Pi_0 (M^{2},T).
\end{eqnarray}
Here, the parameter \( s(T) \) denotes the thermal continuum threshold, which separates the ground state from higher excited states. By equating the Borel-transformed theoretical and hadronic expressions, we derive the thermal current coupling constant sum rule, including contributions up to dimension-four condensates:
\begin{equation}
f^{2}_{T_{4c}}(T)=\frac{e^{m^{2}_{T_{4c}}/M^{2}}}{m^{2}_{T_{4c}}}\Pi (M^{2},T),  \label{eq:Coupl}
\end{equation}
and
\begin{equation}
m^{2}_{T_{4c}}(T)=\frac{\Pi' (M^{2},T)}{\Pi (M^{2},T)},  \label{eq:Mass}
\end{equation}
where $\Pi (M^{2},T)$ is the Borel-transformed and subtracted
invariant amplitude $\Pi ^{ \mathrm{Theor.}}(q^{2})$, \(\Pi' (M^{2},T)\) in Eq.~(\ref{eq:Mass}) denotes the derivative of \(\Pi (M^{2},T)\) with respect to (\(-1/M^{2}\)), i.e., \(\Pi' (M^{2},T) = \frac{d\Pi (M^{2},T)}{d(-1/M^{2})}\).
The subsequent step involves performing a numerical analysis to determine the hadronic parameters of the \( T_{4c} \) resonance. Additionally, by replacing the \( c \) quark with a \( b \) quark, we can obtain the properties of the \( b \)-partner \( T_{4b} \) within the tetraquark framework as a by-product. Note that, due to space constraints, a comprehensive presentation of all temperature-dependent results is not included within the scope of this article. However, for interested readers seeking more detailed information on this topic, the authors are happy to provide the complete data set upon request via email.
%
\section{Analysis and numerical results}
\label{sec:Num}
To extract the mass and current coupling constants for the hidden-charm system \( T_{4c} \) through the TQCDSR approach, several parameters are required, including quark masses, as well as vacuum and thermal gluon condensates. The input data for our calculations include: the charm quark mass (\(m_{c}\)) of \(1.23 \pm 0.09~\mathrm{GeV}\) \cite{Eidemuller:2000rc,ParticleDataGroup:2024cfk}, the bottom quark mass (\(m_{b}\)) of \(4.18^{+0.03}_{-0.02}~\mathrm{GeV}\) \cite{ParticleDataGroup:2024cfk}, and the vacuum gluon condensate value \( \langle 0 \vert G^2 \vert 0 \rangle = 0.028(3)~\mathrm{GeV}^4\) \cite{Horsley:2012ra}. These parameters are crucial for QCD sum rule calculations, providing necessary inputs to evaluate the thermal behavior of the \(T_{4c}\) state.

Additionally, temperature-dependent gluon condensates, as well as energy density, are required. Heavy quarks themselves do not create condensates in the same way light quarks do in QCD. Condensates typically arise from the vacuum expectation values of composite operators involving quark and gluon fields. These composite operators represent combinations of quark and antiquark fields, gluon fields, or their derivatives. 
However, heavy quarks can contribute indirectly to condensates through their interactions with the QCD vacuum. For example, heavy quarks can influence the vacuum structure by inducing polarization effects or by altering the gluon field configurations around them. These effects can lead to modifications in the vacuum condensates, albeit indirectly.

In the context of heavy quark effective theory (HQET), which provides a framework for describing heavy quarks within hadrons, the concept of heavy quark condensates becomes less straightforward. HQET allows for the separation of the dynamics of heavy quarks from that of light quarks and gluons, making it more natural to discuss condensates in terms of light quarks and gluons. In summary, while heavy quarks themselves do not create condensates directly, their presence and interactions can influence the vacuum structure and thereby affect the condensates of light quarks and gluons. Thus we don't take into the heavy quark condensates in our calculations.

The thermal gluon condensates are derived by fitting data from Ref.~\cite{Gubler:2018ctz}, ensuring consistency with Lattice QCD results~\cite{Bazavov:2014pvz,Borsanyi:2013bia}. Eq.~(\ref{G2}) represents the thermal average of the gluon condensate $\langle G^2 \rangle_T$, where $\langle 0 \vert G^2 \vert 0 \rangle$ is the vacuum expectation value of the gluon field strength squared, $C$, $D$, $\beta$, and $\gamma$ are parameters, and $T$ denotes temperature.
\begin{equation}\label{G2}
	\langle G^2 \rangle_T = \langle 0 \vert G^2 \vert 0 \rangle \left[ C + D \left( e^{\beta T - \gamma} + 1 \right)^{-1} \right], 
\end{equation}
where $ C=0.55973 $, $ D = 0.43827 $,  $ \beta = 0.13277~ \mathrm{MeV^{-1}}$, $\mathrm{and}~\gamma = 19.3481 $. 
The below equation represents the expectation value of the gluonic part of the energy-momentum tensor, $\langle \theta_{00}^g \rangle$, as a function of temperature $T$:
\begin{eqnarray}
	\langle \theta_{00}^g \rangle &=& T^4 \exp \left( 113.867 \left[ \frac{1}{\text{GeV}^2} \right] T^2 \right. \notag\\
	& -&  \left. 12.190 \left[ \frac{1}{\text{GeV}} \right] T \right) - 10.141 \left[ \frac{1}{\text{GeV}} \right] T^5. 
\end{eqnarray}
Also, the following expression for the temperature-dependent strong coupling $\alpha_s$ \cite{Kaczmarek:2004gv,Morita:2007hv} is taken into account in the calculations being  $\Lambda_{\overline{MS}}\simeq T_{c}/1.14$ with $ T_c=155 $~MeV~\cite{Andronic:2017pug}:
\begin{eqnarray}\label{geks2T}
	g_{pert}^{2}(T)=\frac{1}{\frac{11}{8\pi^2}\ln\Big(\frac{2\pi T}{\Lambda_{\overline{MS}}}\Big)+\frac{51}{88\pi^2}\ln\Big[2\ln\Big(\frac{2\pi
			T}{\Lambda_{\overline{MS}}}\Big)\Big]}.
\end{eqnarray}
The equation \( g^2 (T)=2.096g_{\text{pert}}^2 (T) \) represents the coupling constant \( g \) at a given temperature \( T \), where \( g_{\text{pert}}^2 (T) \) denotes the perturbative coupling constant. This formula holds for \( T \geq 100 \) MeV. However, for temperatures below \( 100 \) MeV, we substitute the value of \( g^2 (T=100 \text{ MeV}) \).
The below expression describes the temperature dependence of a physical quantity \( s(T) \) in terms of a critical temperature \( T_c \), and other parameters \( s_0 \), \( m_1 \), and \( m_2 \):
\begin{eqnarray}
	s(T) =s_0 \left[ 1 - \left( \frac{T}{T_c} \right)^8 \right] + 4(m_1 + m_2)^2 \left( \frac{T}{T_c} \right)^8,
\end{eqnarray}
where $ s_0 $ is the continuum threshold at zero temperature. At low temperatures (\( T \ll T_c \)), the term \(\left( \frac{T}{T_c} \right)^8\) is very small, so \( s(T) \approx s_0 \). The function \( s(T) \) is dominated by the constant term \( s_0 \).

As the temperature \( T \) approaches the critical temperature \( T_c \), the term \(\left( \frac{T}{T_c} \right)^8\) becomes significant. The first term \( s_0 \left[ 1 - \left( \frac{T}{T_c} \right)^8 \right] \) decreases because \(\left( \frac{T}{T_c} \right)^8 \) approaches 1, reducing the contribution from \( s_0 \). Simultaneously, the second term \( 4 (m_1 + m_2)^2 \left( \frac{T}{T_c} \right)^8 \) increases.

At the critical temperature (\( T = T_c \)), the first term becomes zero because \(\left( \frac{T}{T_c} \right)^8 = 1 \), and \( s(T) \) is dominated by the second term, which is \( 4 (m_1 + m_2)^2 \).

This behavior could be used to model phase transitions or rapid changes in physical properties near a critical temperature, with \( s(T) \) representing quantities like entropy, specific heat, or another order parameter in a thermodynamic system. The parameters \( m_1 \) and \( m_2 \) could relate to masses or other intrinsic properties that affect the system's response to temperature.

The continuum threshold parameter \( s_0\) exhibits some dependence on the mass of the first excited state of \( T_{4c} \). According to QCDSR formalism, physical quantities ideally remain independent of auxiliary parameters like \( M^2 \) and \( s_0 \). Nevertheless, \( M^2 \) and \( s_0 \) are subject to the theory's parameter choices.

In QCDSR methodology, QCD convergence guides us to set a lower bound on \( M^2 \), while the pole contribution (PC) determines an upper limit. Specifically, contributions from the highest-dimensional condensates should ideally not exceed about \( 20\% \) on the theoretical (QCD or OPE) side, and the continuum contribution should be less than \( 50\% \) of the total terms.

\begin{figure}[h!]
	\begin{center}
		\includegraphics[totalheight=6cm,width=8cm]{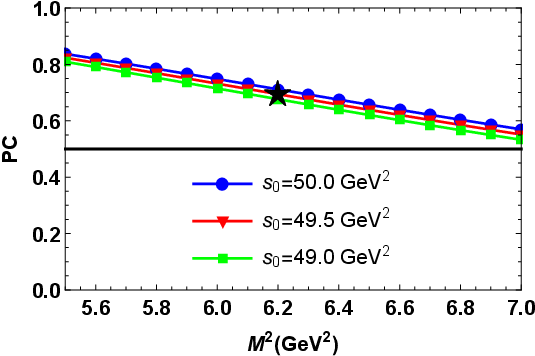}\,\, %
		\includegraphics[totalheight=6cm,width=8cm]{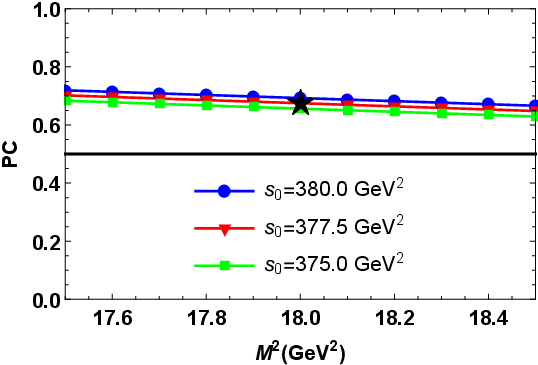}
	\end{center}
	\caption{Pole contribution of the $T_{4c}$ and $T_{4b}$ states, in the tetraquark picture for fixed values of $s_0$ at $ T=0 $, respectively.} \label{PC}
\end{figure}

In this framework, setting \( M^2 \) to its maximum permissible value ensures compliance with the imposed constraint on the pole contribution. This constraint is typical for multiquark systems, and in our study for both states \( T_{4c} \) and \( T_{4b} \):
\begin{equation}
\mathrm{PC}=\frac{\Pi (M^{2},\ s_{0})}{\Pi (M^{2},\ \infty )}=0.68.  \label{eq:PC}
\end{equation}
The lower bound of the Borel window is defined from the convergence of the theoretical side by the ratio below:
\begin{equation}
R_{T_{4c}}(M^{2})=\frac{\Pi ^{\mathrm{(Dim4)}}(M^{2},\ s_{0})}{\Pi (M^{2},\ s_{0})}=0.08,
\label{eq:Convergence1}
\end{equation}
\begin{equation}
R_{T_{4b}}(M^{2})=\frac{\Pi ^{\mathrm{(Dim4)}}(M^{2},\ s_{0})}{\Pi (M^{2},\ s_{0})}=0.01,
	\label{eq:Convergence2}
\end{equation}
here $ \mathrm{Dim4} $ shows the contribution to the
correlation function in the
operator product expansion. Moreover, we present the graphs illustrating the pole contributions of the $T_{4c}$ and $T_{4b}$ states in Figure~\ref{PC}.\\

After considering all these constraints, our analyses lead us to fix the continuum threshold and Borel parameters for the \( T_{4c} \) resonance and its \( b \)-partner \( T_{4b} \). Our results are shown in Table~\ref{tab:M2s0}:
\begin{table}[h!]
	\centering
	\begin{tabular}{|c|c|c|}
		\hline
		\textbf{Resonance} & \textbf{\( M^2 \) (GeV\(^2\))} & \textbf{\( s_{0} \) (GeV\(^2\))} \\ \hline
		\( T_{4c} \) & \( 5.5-7.0 \) & \( 49.0-50.0 \) \\ \hline
		\( T_{4b} \) & \( 17.5-18.5 \) & \( 375.0-380.0 \) \\ \hline
	\end{tabular}
	\caption{Specified intervals of Borel mass parameters and continuum thresholds for \( T_{4c} \) and \( T_{4b} \) resonances.}
	\label{tab:M2s0}
\end{table}

By the principles of the QCDSR method, the consistency of hadronic quantities concerning the Borel parameter \( M^2 \) and continuum threshold \( s_0 \) within the chosen operational range ensures the reliability of results derived from the sum rules.

\begin{table}[h!]
	\centering
	\begin{tabular}{|c|c|c|}
		\hline
		\textbf{Resonance} & \textbf{Mass (MeV)} & \textbf{Current coupling (GeV\(^{4}\))} \\ \hline
		\( T_{4c} \) & \( (6552.70 \pm 50.04) \) & \( (5.16 \pm 0.35) \times 10^{-2} \) \\ \hline
		\( T_{4b} \) & \( (18499.60 \pm 44.55) \) & \( (6.54 \pm 0.38) \times 10^{-1} \) \\ \hline
	\end{tabular}
	\caption{Mass and current coupling constant of \( T_{4c} \) and \( T_{4b} \) resonances in the \( T=0 \) value.}
	\label{tab:M2s0_T0}
\end{table}

\begin{table}[h!]
	\centering
	\begin{tabular}{|c|c|c|}
		\hline
		\textbf{Resonance} & \textbf{Mass (MeV)} & \textbf{Current coupling (GeV\(^{4}\))} \\ \hline
		\( T_{4c} \) & \( (6027.60 \pm 21.33) \) & \( (1.49 \pm 0.08) \times 10^{-2} \) \\ \hline
		\( T_{4b} \) & \( (17885.60 \pm 28.66) \) & \( (2.18 \pm 0.12) \times 10^{-1} \) \\ \hline
	\end{tabular}
	\caption{Mass and current coupling constant of \( T_{4c} \) and \( T_{4b} \) resonances in the \( T=0.14~\mathrm{GeV} \) value.}
	\label{tab:M2s0_T014}
\end{table}

These findings are consistent with both experimental data and theoretical estimations, considering the uncertainties involved.

\begin{figure}[h!]
	\begin{center}
		\includegraphics[totalheight=6cm,width=8cm]{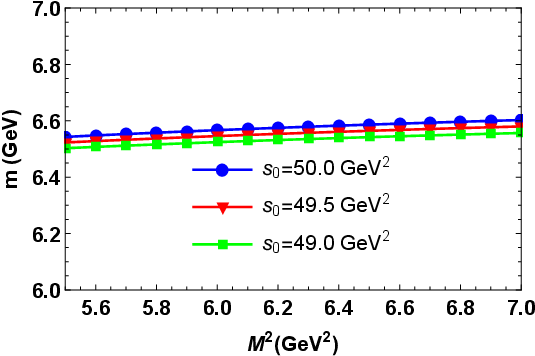}\,\, %
		\includegraphics[totalheight=6cm,width=8cm]{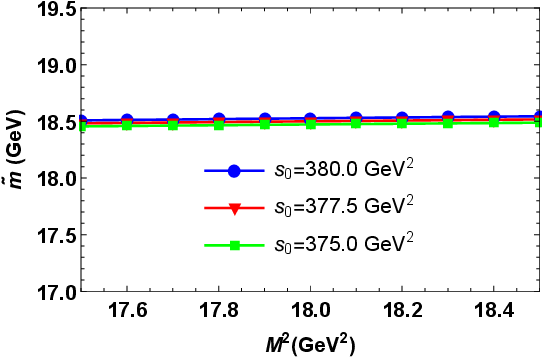}
	\end{center}
	\caption{Vacuum masses of \( T_{4c} \) and \( T_{4b} \) states versus the Borel mass parameter in the tetraquark picture, for fixed values of \( s_0 \) at $ T=0 $, respectively.} \label{massvsM2}
\end{figure}

\begin{figure}[h!]
	\begin{center}
		\includegraphics[totalheight=6cm,width=8cm]{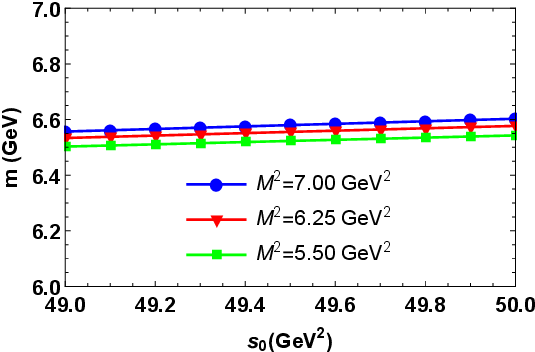}\,\, %
		\includegraphics[totalheight=6cm,width=8cm]{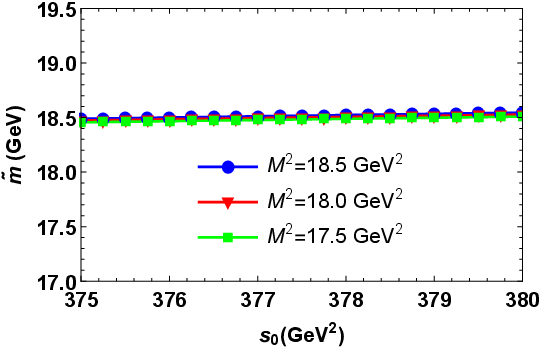}
	\end{center}
	\caption{Vacuum masses of \( T_{4c} \) and \( T_{4b} \) states versus the continuum threshold in the tetraquark model, using fixed \( M^2 \) values.} \label{massvss0}
\end{figure}

\begin{figure}[h!]
	\begin{center}
		\includegraphics[totalheight=6cm,width=8cm]{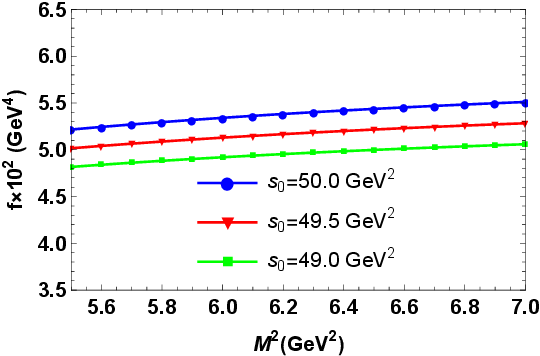}\,\, %
		\includegraphics[totalheight=6cm,width=8cm]{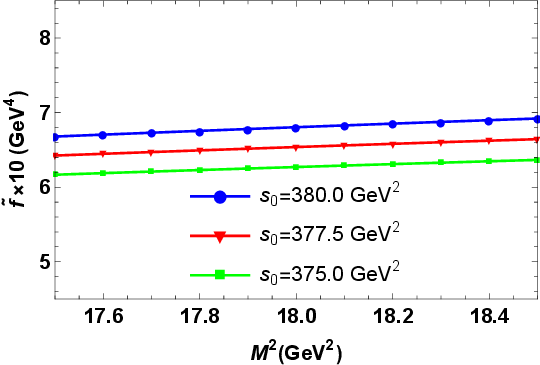}
	\end{center}
	\caption{The current coupling constants of \( T_{4c} \) and \( T_{4b} \) states versus Borel mass parameter in the tetraquark picture, for fixed values of \( s_0 \), respectively.} \label{fvsM2}
\end{figure}

\begin{figure}[h!]
	\begin{center}
		\includegraphics[totalheight=6cm,width=8cm]{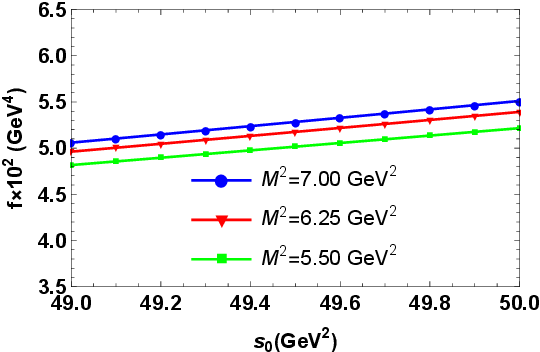}\,\, %
		\includegraphics[totalheight=6cm,width=8cm]{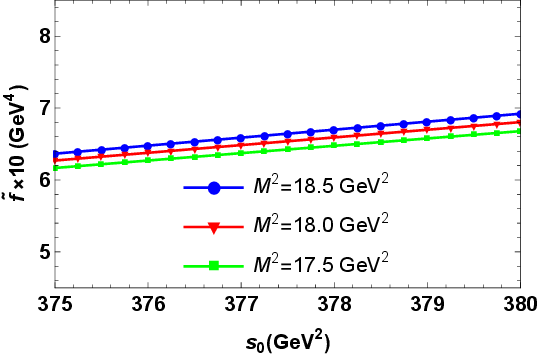}
	\end{center}
	\caption{Continuum threshold dependence of the current coupling constants for \( T_{4c} \) and \( T_{4b} \) states within the tetraquark model, with fixed \( M^2 \) values.} \label{fvss0}
\end{figure}

We see the stability of sum rules according to the model parameters drawing graphs. Here we present plots for the $T_{4c}$ and  $T_{4b}$ states in Figure~\ref{massvsM2}, and \ref{massvss0}. Also, we draw these states' current coupling constants versus the Borel parameter $M^2$ and $s_0$  in Figures~\ref{fvsM2}, and \ref{fvss0}, respectively. 

\begin{figure}[h!]
	\begin{center}
		\includegraphics[totalheight=6cm,width=8cm]{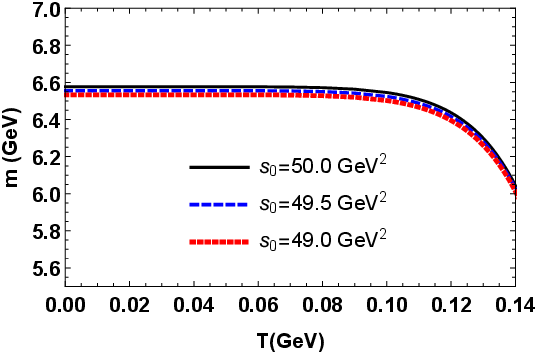}\,\, %
		\includegraphics[totalheight=6cm,width=8cm]{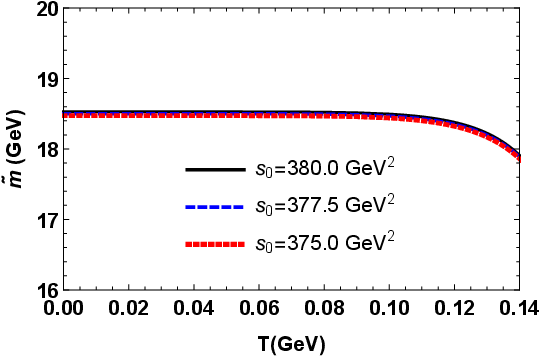}
	\end{center}
	\caption{Temperature dependence of the masses of \( T_{4c} \) and \( T_{4b} \) states in the tetraquark picture, for fixed values of \( s_0 \), respectively.} \label{massvsT}
\end{figure}

\begin{figure}[h!]
	\begin{center}
		\includegraphics[totalheight=6cm,width=8cm]{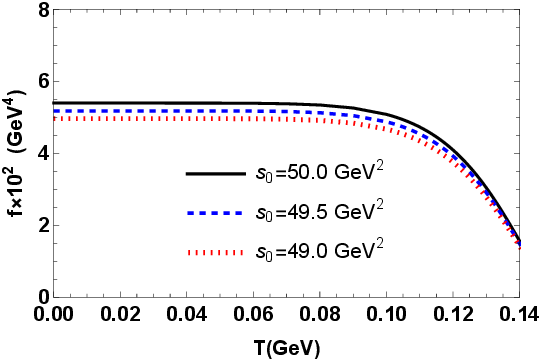}\,\, %
		\includegraphics[totalheight=6cm,width=8cm]{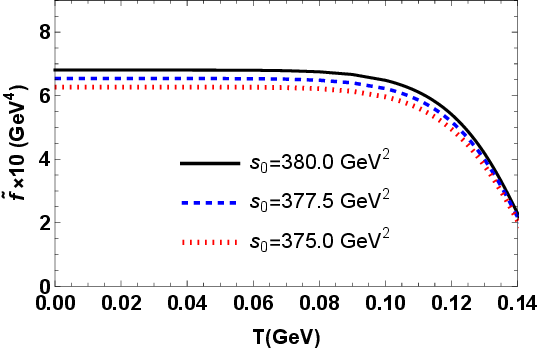}
	\end{center}
	\caption{The variation of the current coupling constant with temperature for \( T_{4c} \) and \( T_{4b} \) states in the tetraquark model, using fixed \( s_0 \) values.} \label{fvsT}
\end{figure}

We now explore how the mass and current coupling constant of \( T_{4c} \) and \( T_{4b} \) vary with temperature, as illustrated by the graphs in Figures~\ref{massvsT} and \ref{fvsT} for the tetraquark model.

\clearpage
\section{\textbf{Conclusion}}\label{result}
%
In this study, the TQCDSR approach was employed to investigate the spectroscopic parameters of the fully heavy tetraquarks \( T_{4c} \) and \( T_{4b} \) under varying temperature conditions. Our numerical analyses yielded insightful results on the behavior of these tetraquark states as the temperature increased.

These findings indicate a significant decrease in both masses and coupling constants with increasing temperature. The mass of the \( T_{4c} \) state has dropped to approximately $ 92\% $ of its vacuum value, while the mass of \( T_{4b} \) has decreased to about $ 96.7\% $ of the vacuum value. Current coupling constants have also experienced marked reductions; the \( T_{4c} \) state has diminished to approximately 29\% of the vacuum value, whereas the \( T_{4b} \) state has retreated to about $ 33.4\% $ of the vacuum value. These ratios demonstrate that increasing temperature has had significant effects on the internal structure and binding properties of tetraquarks.

These results indicate that the masses and current coupling constants of the fully heavy tetraquarks \( T_{4c} \) and \( T_{4b} \) are sensitive to temperature changes. Further experimental validation, particularly in high-energy collisions and future particle physics experiments, is essential to corroborate these findings and explore their implications for QCD and the study of exotic hadronic matter.
\appendix*

\section{ The spectral density }

\renewcommand{\theequation}{\Alph{section}.\arabic{equation}} \label{sec:App}

This appendix contains the spectral density, which
has been used to compute the mass of the scalar fully-charm and bottom tetraquarks. The components of the spectral density $\rho ^{\mathrm{Theor.}}(s,T)$ are given by the general expression:
\begin{widetext}

\begin{eqnarray}
	\rho ^{\mathrm{Theor.}}(s,T)=\int_{0}^{1}dr \int_{0}^{1-r }dw \int_{0}^{1-r-w }dz  \rho (s,T,r ,w ,z ),  \label{eq:A4}
\end{eqnarray}
where $r $, $w $, and $z $ are the Feynman parameters. Here, $\rho(s, T, r, w, z) = \rho^{\text{pert}}(s, T, r, w, z) + \rho^{\text{Dim4}}(s, T, r, w, z)$.

The perturbative function $\rho ^{\mathrm{pert.}}(s, T,  r,  w,  z)$
is given by the formula:
\begin{eqnarray}
	\rho ^{\mathrm{pert.}}(s, T,  r,  w,  z)&=&\frac{\Theta ({N)}(E m_Q^2-A  s r w z)^{2}}{64 \pi^6 E^2}\left\{-21 A^3 s^2 r^3 w^3 z^3 + A E m_Q^2 s r w z\left[ -5 C r^2 w^2 +  z(13 r w (r+B)
	\right. \right.    \notag \\
	&+&\left. 5z(B^2 w+B r(3w-1)+r^2 (r+3 w-2))+5 r z^2 (2 r +B))\right]+E^2 m_Q^4 \left[ 2 C r^2 w^2 
	\right. \notag \\
	&\times &
	(3 r+3 w-2)+z(2 B^2 w z(3w-1)+B r
	(12w^3-12w^2-w+2z+6wz(4w-3)
	\notag \\  
	&+&2z^2(6w-1))+r^2(36w^3+12w^2(3r+5z-4)+2z(5-7r+3r^2-8z+6rz+3z^2))
	\notag \\ 
	&+&\left. \left.w(11-54z-24r+12r^2+36rz+36z^2))
	\right]
	\right\} .
\end{eqnarray}
The nonperturbative function $\rho ^{\mathrm{Dim4}}(s, T,  r,  w,  z)$
is given by the formula:
\begin{eqnarray}
	\rho ^{\mathrm{Dim4}}(s, T,  r,  w,  z)&=&\langle \frac{\alpha_{s}GG}{\pi}\rangle_{T} \frac{\Theta (N)}{192 \pi^4 E^6}\left\{-3 C^2 E m_Q^2 \left[s r w z (r^2w^2 (r^2+w^2-3A(r+w))+3 A r w z (3w+3r+4rw)
	\right. \right. 
	\notag \\ 
	&+& 
	z^4(-11r^2+11w-11w^2+11r+2wr) )+E m_Q^2 (2 r^2 w^2 (r+w)^2+r w z (4r^3+4r^2(3w-1)
	\notag \\ 
	&+& \left.
	w(4w^2-4w-1)+r(12w^2-16w-1))-4z^5 (r+w))
	\right]+C \left[ 30 A^3 s^2 r^3 w^3 z^3 (r+w)
	\right. 
	\notag \\ 
	&+& 
	3 E m_Q^2 s r w z^2 (r w (-3+3r(4-6r+3r^2))+12w+r w (35 r -9r^2 -36)+w^2(35r-16r^2-18)
	\notag \\ 
	&-& 
	9w^3(r-1) )+z (r^5-4r^4-3Ar^4+3wr^4+r+w(w-1)^2(1+w^2-3Aw-2w)
	\notag \\ 
	&+& 
	r^3(6+26w+6A-24Aw-30w^2)+r w (6-3w^3+26w^2-24w)-3A(2-7w+8w^2)
	\notag \\ 
	&-& 
	r^2(4+30w^3-60w^2+24w)+3A(1-7w+18w^2))-6A(r+w)z^2(r^2+w^2-w-r+6rw)
	\notag \\ 
	&+& 
	z^4(13r^2+13w^2-13w+18rw-13r)+8z^5(r+w)+E^2m_Q^4 (r^2 w^2 (-1+3r^3+r^2(-3+5w)
	\notag \\ 
	&+& 
	3w(1-w+w^2)+r(3+5w^2-6w))+rwz(-4+6r^4+4r^3(5w-3)+3w(w-1)(2w^2-2w-3)
	\notag \\ 
	&+&
	r^2(-3+28w^2-24w)+9r-3 r w-24w^2 r+20r w^3)-3z^2(2r^5+2w^2(w-1)^3+2r^4(7w-3)
	\notag \\ 
	&+&
	r^3(6-36w+32w^2)+rw(-4+25w+14w^3-36w^2)+r^2(-2+25w+32w^3-68w^2)))
	\notag \\ 
	&-& \left.
	2z^3(r^4+4r-rw-28rw^2+4w-6w^2+w^4-6r^2-28r^2w-2r^2w^2)+4z^6 (r+w)
	\right]
	\notag \\ 
	&+& E m_Q^2 z^2 \left[3 s r w z^2 (5r^5+r^4(w-20)+r^3(w-30w^2-3 A z +30)+w(5(w-1)^4-3 Aw z (w-1)
	\right.
	\notag \\ 
	&+&
	+2z^4)+r^2(3 A z-20+3 w (3+14w-10w^2-7A z))+r(5+w^3+w^4+2z^4+w^2(9-21Az)
	\notag \\ 
	&+&
	+w(6 A z-13)))+E m_Q^2(3r^6+4r^5(7w-3)+w(w-1)^2(+6w-6w^2+3w^3-2)+wz^2
	\notag \\ 
	&+&
	(-14+36w-30w^2+9w^3)+r^4(21-76w+81w^2+9z^2)+2r^3(-10+33w-78w^2+56w^3
	\notag \\ 
	&+&
	-15z^2+20wz^2)
	+r(-2+5w-25w^2+66w^3-76w^4+28w^5-14z^2+61z^2w-62z^2w^2
	\notag \\ 
	&+&\left.\left.
	+40z^2w^3)+r^2(10+36z^2+w(-25+93w-156w^2+81w^3+62z^2(w-1))))
	\right]
	\right\} 
	\notag \\ 
	&+&g^{2}(T) \langle u^{\lambda}\theta_{\lambda\sigma}^{g}u^{\sigma}\rangle_{T}\frac{\Theta (N)}{64 \pi^6 E^6} A s r w z \left\{5 A^2 C s r^2 w^2 z^2 (r+w)+m_Q^2(C r w+c z(r+w)+z^2(r+w))
	\right.
	\notag \\ 
	&\times&
	\left[C r^2 w^2(r^2+w(w-1)+r(6w-1))-C r w z(3r^2+3w(w-1)+r(2w-3))+z^3(r^4(3-5w)
	\right.
	\notag \\ 
	&-&
	r^5-w^2(w-1)^3+r^3(5w-10w^2-3)-r w(w-1)(5w^2-2)+r^2(1+2w+8w^2-10w^3))
	\notag \\ 
	&+&
	+2 C z^3(r^3+3r^2(w-1)+w(w-1)(w-2)+r(2+3w^2-10w))+z^4(11r^3+w(w-1)
	\notag \\ 
	&+&
	\left.\left.
	(11w-12)+r^2(33w-23)+r(12-50w+33w^2))+12Cz^5(r+w)+4z^6(r+w)
	\right]
	\right\} 
\end{eqnarray}
where $\Theta ({N)}$ is the unit step function.  

The nonperturbative contribution $ \Pi_0 (M^{2},T) $ function is given as:
\begin{eqnarray}
	\Pi_0 (M^{2},T)&=&\langle \frac{\alpha_{s}GG}{\pi}\rangle_{T} \frac{m_Q^6}{384 \pi^4} \int_{0}^{1}dr \int_{0}^{1-r }dw \int_{0}^{1-r-w }dz \frac{1}{A E^3 r w z}\left\{B^2 w^4 z^2 +r^6 (w+z)^2+2 B r^5 (w+z)(2w+z)	\right.	\notag \\ 
	&+&B r w z(-2+w(w-1)
	(2w^2-2w-5)+5z+3wz(2w^2-3)+(4w^2+4w-5))+r^4(6w^2(w-1)^2+w z
	\notag \\ 
	&\times&
	(20w^2-20w+1)+z^2(23w^2-10w+1)+2z^3(5w-1)+z^4)+r^2 w (w(w-1)^4+z(-7+5w(3+2w^2
	\notag \\ 
	&\times&
	(w-2)))+z^2(19++w(-25+w(23w-16)))+2z^3(10w^2+2w-9)+2z^4(3w+2))+4B r^3 w(w^3+w
	\notag \\ 
	&+&
	\left.
	w^2(4z-2)+w z(4z-1)+z(z^2+z-2))\right\}  Exp \left[ -\frac{m_Q^2 F}{M^2}\right]
\end{eqnarray}
where
\begin{eqnarray}
	N&=&\dfrac{s r w z(r^3+B^2 z+B r(2r+2z+w-1))-Em_Q^2(r^2+B(r+z))}{E^2},
	\notag \\ 
	F&=&\dfrac{r w z-A (wz +rz+rw)}{Arwz},
	\notag \\ 
	E&=&r^2(w+z)+B(rw+rz+wz),
	\notag \\ 
	C&=&r+w-1,
	\notag \\ 
	B&=&z+w-1,
	\notag \\ 
	A&=&z+r+w-1.
\end{eqnarray}
\end{widetext}


\begin{thebibliography}{99}
	
\bibitem{Shuryak:2008eq}
E.~Shuryak,
\href{https://www.sciencedirect.com/science/article/pii/S0146641008000732?via}{Prog. Part. Nucl. Phys. \textbf{62}, 48-101 (2009)}

\bibitem{Ayala:2020rmb}
A.~Ayala, S.~Hernandez-Ortiz, L.~A.~Hernandez, V.~Knapp-Perez and R.~Zamora,
\href{https://journals.aps.org/prd/abstract/10.1103/PhysRevD.101.074023}{Phys. Rev. D \textbf{101}, no.7, 074023 (2020)}

\bibitem{Ayala:2016vnt}
A.~Ayala, C.~A.~Dominguez and M.~Loewe,
\href{https://www.hindawi.com/journals/ahep/2017/9291623/}{Adv. High Energy Phys. \textbf{2017}, 9291623 (2017)}

\bibitem{Steinbrecher:2018phh}
Steinbrecher, P. [HotQCD Collaboration],
\href{https://doi.org/10.1016/j.nuclphysa.2018.08.02}{Nucl.\ Phys. \ A {\bf 982}, 847 (2019)}

\bibitem{Fischer:2018sdj}
C.~S.~Fischer,
\href{https://doi.org/10.1016/j.ppnp.2019.01.002}{Prog. Part. Nucl. Phys. \textbf{105},  1-60 (2019)}
	
\bibitem{ALICE:2008ngc}
K.~Aamodt \textit{et al.} [ALICE],
\href{https://iopscience.iop.org/article/10.1088/1748-0221/3/08/S08002}{JINST \textbf{3}, S08002 (2008)}
	
\bibitem{STAR:2005gfr}
J.~Adams \textit{et al.} [STAR],
\href{https://doi.org/10.1016/j.nuclphysa.2005.03.085}{Nucl. Phys. A \textbf{757}, 102-183 (2005)}

\bibitem{CMS:2024krd}
A.~Hayrapetyan \textit{et al.} [CMS],
\href{https://arxiv.org/abs/2405.10785}{[arXiv:2405.10785 [nucl-ex]]}

\bibitem{STAR:2024bpc}
STAR Collaboration
\href{https://arxiv.org/abs/2402.01998}{[arXiv:2402.01998 [nucl-ex]]}

\bibitem{Wang:2023xhn}
C.~Z.~Wang [ALICE],
\href{https://www.epj-conferences.org/articles/epjconf/abs/2024/06/epjconf_QuarkMatter2023_04007/epjconf_QuarkMatter2023_04007.html}{EPJ Web Conf. \textbf{296}, 04007 (2024)}

\bibitem{Podlaski:2024kxg}
P.~Podlaski,
\href{https://www.epj-conferences.org/articles/epjconf/abs/2024/06/epjconf_QuarkMatter2023_01008/epjconf_QuarkMatter2023_01008.html}{EPJ Web Conf. \textbf{296}, 01008 (2024)}
	
\bibitem{Brambilla:2019esw}
N.~Brambilla, S.~Eidelman, C.~Hanhart, A.~Nefediev, C.~P.~Shen, C.~E.~Thomas, A.~Vairo and C.~Z.~Yuan,
\href{https://doi.org/10.1016/j.physrep.2020.05.001}{Phys. Rept. \textbf{873}, 1-154 (2020)}

\bibitem{Chen:2016qju}
H.~X.~Chen, W.~Chen, X.~Liu and S.~L.~Zhu,
\href{https://doi.org/10.1016/j.physrep.2016.05.004}{Phys. Rept. \textbf{639}, 1-121	 (2016)}

\bibitem{Agaev:TJP}
S. S. Agaev, K. Azizi and H. Sundu, 
\href{https://journals.tubitak.gov.tr/physics/vol44/iss2/1/}{Turk. J. Phys. \textbf{44}, no. 2, 95 (2020)}

\bibitem{Azizi:2020itk}
K.~Azizi and N.~Er,
\href{https://doi.org/10.1103/PhysRevD.101.074037}{Phys. Rev. D \textbf{101}, no.7, 074037 (2020)}

\bibitem{Azizi:2020yhs}
K.~Azizi and N.~Er,
\href{https://doi.org/10.1016/j.physletb.2020.135979}{Phys. Lett. B \textbf{811}, 135979 (2020)}

\bibitem{Sungu:2024oax}
J.~Y.~S\"ung\"u and N.~Er,
\href{https://iopscience.iop.org/article/10.1088/1361-6471/ad66eb}{J. Phys. G \textbf{51}, no.12, 125001 (2024)}

\bibitem{Veliev:2014tca}
E.~V.~Veliev, K.~Azizi, H.~Sundu and G.~Kaya,
\href{https://rjp.nipne.ro/2014_59_1-2/RomJPhys.59.p140.pdf}{Rom. J. Phys. \textbf{59-63}, no.1-2, 140 (2014)}

\bibitem{Veliev:2011kq}
E.~V.~Veliev, K.~Azizi, H.~Sundu, G.~Kaya and A.~T\"urkan,
\href{https://link.springer.com/article/10.1140/epja/i2011-11110-8}{Eur. Phys. J. A \textbf{47}, 110 (2011)}

\bibitem{Sungu:2020zvk}
J.~Y.~S\"ung\"u, A.~T\"urkan, H.~Sundu and E.~V.~Veliev,
\href{https://link.springer.com/article/10.1140/epjc/s10052-022-10305-0}{Eur. Phys. J. C \textbf{82}, no.5, 453 (2022)}

\bibitem{Sungu:2019ybf}
J.~Y.~S\"ung\"u, A.~T\"urkan and E.~Veli Veliev,
\href{https://www.actaphys.uj.edu.pl/index_n.php?I=R&V=50&N=9#1501}{Acta Phys. Polon. B \textbf{50}, 1501 (2019)}

\bibitem{Turkan:2019anj}
A.~T\"urkan, H.~Da\u{g}, J.~Y.~S\"ung\"u and E.~Veli Veliev,
\href{https://iopscience.iop.org/article/10.1209/0295-5075/126/51001}{EPL \textbf{126}, no.5, 51001 (2019)}

\bibitem{Gungor:2023ksu}
E.~G\"ung\"or, H.~Sundu, J.~Y.~S\"ung\"u and E.~V.~Veliev,
\href{https://link.springer.com/article/10.1007/s00601-023-01807-y}{Few Body Syst. \textbf{64}, no.3, 53 (2023)}

\bibitem{Sungu:2020azn}
J.~Y.~S\"ung\"u, A.~T\"urkan, E.~Sertbakan and E.~V.~Veliev,
\href{https://link.springer.com/article/10.1140/epjc/s10052-020-08439-0}{Eur. Phys. J. C \textbf{80}, no.10, 943 (2020)}

\bibitem{liu:2020eha}
M.~S.~Liu, F.~X.~Liu, X.~H.~Zhong and Q.~Zhao,
\href{https://journals.aps.org/prd/abstract/10.1103/PhysRevD.109.076017}{Phys. Rev. D \textbf{109}, no.7, 076017 (2024)}

\bibitem{LHCb:2020bwg}
R.~Aaij \textit{et al.} [LHCb],
\href{https://doi.org/10.1016/j.scib.2020.08.032}{Sci. Bull. \textbf{65}, no.23, 1983-1993 (2020)}

\bibitem{CMS:2023owd}
A.~Hayrapetyan \textit{et al.} [CMS],
\href{https://journals.aps.org/prl/abstract/10.1103/PhysRevLett.132.111901}{Phys. Rev. Lett. \textbf{132}, no.11, 111901 (2024)}

\bibitem{Agaev:2023wua}
S.~S.~Agaev, K.~Azizi, B.~Barsbay and H.~Sundu,
\href{https://www.sciencedirect.com/science/article/pii/S0370269323004239?via%3Dihub}{Phys. Lett. B \textbf{844}, 138089 (2023)}
		
	\bibitem{Faustov:2022mvs}
	R.~N.~Faustov, V.~O.~Galkin and E.~M.~Savchenko,
	\href{https://www.mdpi.com/2073-8994/14/12/2504}{Symmetry \textbf{14}, no.12, 2504 (2022)}
	
	\bibitem{Lin:2024olg}
	Y.~Y.~Lin, J.~Y.~Wang and A.~Zhang,
	\href{https://link.springer.com/article/10.1140/epjp/s13360-024-05480-w}{Eur. Phys. J. Plus \textbf{139}, no.8, 707 (2024)}

	\bibitem{Chen:2016jxd}
	W.~Chen, H.~X.~Chen, X.~Liu, T.~G.~Steele and S.~L.~Zhu,
	\href{https://www.sciencedirect.com/science/article/pii/S0370269317306561?via}{Phys. Lett. B \textbf{773}, 247-251 (2017)}

		\bibitem{Wang:2022xja}
		Z.~G.~Wang,
		\href{https://doi.org/10.1016/j.nuclphysb.2022.115983}{Nucl. Phys. B \textbf{985}, 115983 (2022)}
				
		\bibitem{Anwar:2023fbp}
		M.~N.~Anwar and T.~J.~Burns,
		\href{https://journals.aps.org/prd/abstract/10.1103/PhysRevD.110.034012}{Phys. Rev. D \textbf{110}, no.3, 034012 (2024)}
		
		\bibitem{Yan:2023lvm}
		T.~Q.~Yan, W.~X.~Zhang and D.~Jia,
		\href{https://link.springer.com/article/10.1140/epjc/s10052-023-11956-3}{Eur. Phys. J. C \textbf{83}, no.9, 810 (2023)}
		
		\bibitem{Yang:2020wkh}
		B.~C.~Yang, L.~Tang and C.~F.~Qiao,
		\href{https://link.springer.com/article/10.1140/epjc/s10052-021-09096-7}{Eur. Phys. J. C \textbf{81}, no.4, 324 (2021)}
	
		\bibitem{Yu:2022lak}
		G.~L.~Yu, Z.~Y.~Li, Z.~G.~Wang, J.~Lu and M.~Yan,
		\href{https://link.springer.com/article/10.1140/epjc/s10052-023-11445-7}{Eur. Phys. J. C \textbf{83}, no.5, 416 (2023)}
	
\bibitem{Brambilla:2010cs}
N.~Brambilla, S.~Eidelman, B.~K.~Heltsley, R.~Vogt, G.~T.~Bodwin, E.~Eichten, A.~D.~Frawley, A.~B.~Meyer, R.~E.~Mitchell and V.~Papadimitriou, \textit{et al.}
\href{https://link.springer.com/article/10.1140/epjc/s10052-010-1534-9}{Eur. Phys. J. C \textbf{71}, 1534 (2011)},

\bibitem{Shifman:1978bx}  M. A. Shifman, A. I. Vainshtein and V. I. Zakharov,
\href{https://doi.org/10.1016/0550-3213(79)90022-1}{Nucl. Phys. B \textbf{147}, 385 (1979)}

\bibitem{Reinders:1984sr} 
L.~J.~Reinders, H.~Rubinstein and S.~Yazaki,
\href{https://doi.org/10.1016/0370-1573(85)90065-1}{Phys.\ Rept.\  {\bf 127}, 1 (1985)}

\bibitem{Bochkarev:1985ex}
A.~I.~Bochkarev and M.~E.~Shaposhnikov,
\href{https://doi.org/10.1016/0550-3213(86)90209-9}{Nucl.\ Phys.\ B {\bf 268}, 220 (1986)}

\bibitem{Azizi:2019cmj}
K.~Azizi and A.~T\"{u}rkan,
\href{https://link.springer.com/article/10.1140/epjc/s10052-020-7931-9}{Eur. Phys. J. C \textbf{80}, no.5, 425 (2020)}

\bibitem{Mallik:1997pq}
S.~Mallik,
\href{https://doi.org/10.1016/S0370-2693(97)01335-X}{Phys. Lett. B \textbf{416}, 373-378 (1998)}

\bibitem{Furnstahl:1992pi}
R.~J.~Furnstahl, D.~K.~Griegel and T.~D.~Cohen,
\href{https://journals.aps.org/prc/abstract/10.1103/PhysRevC.46.1507}{Phys. Rev. C \textbf{46}, 1507-1527 (1992)}

\bibitem{Dominguez:2010mx}
C.~A.~Dominguez, M.~Loewe, J.~C.~Rojas and Y.~Zhang,
\href{https://journals.aps.org/prd/abstract/10.1103/PhysRevD.83.034033}{Phys. Rev. D \textbf{83}, 034033 (2011)}

\bibitem{Gubler:2011ua}
P.~Gubler, K.~Morita and M.~Oka,
\href{https://journals.aps.org/prl/abstract/10.1103/PhysRevLett.107.092003}{Phys. Rev. Lett. \textbf{107}, 092003 (2011)}

\bibitem{Azizi:2016ddw}
K.~Azizi and G.~Bozk\i r,
\href{https://link.springer.com/article/10.1140/epjc/s10052-016-4370-8}{Eur. Phys. J. C \textbf{76}, no.10, 521 (2016)}

\bibitem{Eidemuller:2000rc}
M.~Eidemuller and M.~Jamin,
\href{https://doi.org/10.1016/S0370-2693(00)01391-5}{Phys. Lett. B \textbf{498}, 203-210 (2001)}

\bibitem{ParticleDataGroup:2024cfk}
S.~Navas \textit{et al.} [Particle Data Group],
\href{https://journals.aps.org/prd/abstract/10.1103/PhysRevD.110.030001}{Phys. Rev. D \textbf{110}, no.3, 030001 (2024)}

\bibitem{Horsley:2012ra}
R.~Horsley, G.~Hotzel, E.~M.~Ilgenfritz, R.~Millo, H.~Perlt, P.~E.~L.~Rakow, Y.~Nakamura, G.~Schierholz and A.~Schiller,
\href{https://journals.aps.org/prd/abstract/10.1103/PhysRevD.86.054502}{Phys. Rev. D \textbf{86}, 054502 (2012)}

\bibitem{Gubler:2018ctz}
P.~Gubler and D.~Satow,
\href{Gubler:2018ctz}{Prog.\ Part.\ Nucl.\ Phys.\  {\bf 106}, 1 (2019)}

\bibitem{Bazavov:2014pvz}
A.~Bazavov \textit{et al.} [HotQCD],
\href{https://journals.aps.org/prd/abstract/10.1103/PhysRevD.90.094503}{Phys. Rev. D \textbf{90}, 094503 (2014)}

\bibitem{Borsanyi:2013bia}
S.~Borsanyi, Z.~Fodor, C.~Hoelbling, S.~D.~Katz, S.~Krieg and K.~K.~Szabo,
\href{https://doi.org/10.1016/j.physletb.2014.01.007}{Phys. Lett. B \textbf{730}, 99-104 (2014)}

\bibitem{Kaczmarek:2004gv}
O.~Kaczmarek, F.~Karsch, F.~Zantow and P.~Petreczky,
\href{https://journals.aps.org/prd/abstract/10.1103/PhysRevD.70.074505}{Phys. Rev. D \textbf{70}, 074505 (2004)
[erratum: Phys. Rev. D \textbf{72}, 059903 (2005)]}

\bibitem{Morita:2007hv}
K.~Morita and S.~H.~Lee,
\href{https://journals.aps.org/prc/abstract/10.1103/PhysRevC.77.064904}{Phys. Rev. C \textbf{77}, 064904 (2008)}

\bibitem{Andronic:2017pug}
A.~Andronic, P.~Braun-Munzinger, K.~Redlich and J.~Stachel,
\href{https://www.nature.com/articles/s41586-018-0491-6}{Nature {\bf 561}, no. 7723, 321 (2018)}

\end{thebibliography}
\end{document}